\documentclass[aip,author-numerical,jmp,showpacs]{revtex4-1}
\pdfoutput=1
\usepackage[sumlimits]{amsmath}
\usepackage{latexsym}
\usepackage{amssymb}
\usepackage{euscript}
\usepackage{graphicx}

\begin{document}
\tolerance=5000

\author{Gustavo Arciniega}
\email{gustavo.arciniega, fnettel, leopj@ciencias.unam.mx}
\author{Francisco Nettel}
\email{gustavo.arciniega, fnettel, leopj@ciencias.unam.mx} 
\author{Leonardo Pati\~no}
\email{gustavo.arciniega, fnettel, leopj@ciencias.unam.mx}
\affiliation{Departamento de F\'\i sica, Facultad de Ciencias, Universidad Nacional Aut\'onoma de M\'exico, \\
A.P. 50-542, M\'exico D.F. 04510, M\'exico}
\author{Hernando Quevedo}
\email{quevedo@nucleares.unam.mx} 
\affiliation{Instituto de Ciencias Nucleares, Universidad Nacional Aut\'onoma de M\'exico, \\
A.P. 70-543, M\'exico D.F. 04510, M\'exico}

\date{\today}

\begin{abstract}
We present a topological quantization of free massive bosonic fields as the first example of a classical field theory with a quantum counterpart to be studied under this formalism. First, we identify certain harmonic map as a geometric representation of this physical system. We take as a concrete example the case of free massive bosonic fields in two dimensions represented by the minimal embedding of a two dimensional surface into a pp-wave spacetime. We use this geometric representation to construct the fiber bundle corresponding to some specific field configurations and then find their topological spectra, defined in terms of the Euler invariant. 
We discuss the results for some particular configurations and their consequences for the energy of the system. 
\end{abstract}

\maketitle

\section{Introduction}

The formalism of topological quantization in the way we will understand it here was formulated in a series of articles where a detailed description of the method was presented \cite{Patino:2002jp,Patino:2004wc,Nettel:2007zz,Nettel03}. In general, topological quantization procedures are inspired by Dirac's seminal paper \cite{Dirac:1931kp} about electric-magnetic charge quantization and its topological formulation given years later \cite{WuYang1975}. In the topological context, the charge quantization is a consequence of the regularity condition that must be obeyed by the potential connection $A_{\mu}$ in the principal fiber bundle (pfb). 

Since Dirac's quantization of the electric-magnetic charge, there have been attempts to find quantization conditions for other physical systems. Pati\~no and Quevedo \cite{Patino:2004wc} formulated a procedure to determine discrete relationships among the parameters that enter in the description of gravitational fields. These and other physical systems \cite{Nettel:2007zz,Nettel03,Arciniega0} were studied with this method, which is basically the construction of a pfb which represents such systems and the analysis of its topological characteristics. 

Pati\~no and Quevedo distinguished between two kinds of such procedures and referred to them as a) induced topological quantization and b) intrinsic topological quantization.

The induced topological quantization take its name from the fact that the fiber bundle to be analyzed and its topological characteristics are dictated by a physical gauge field living on the base space, so its standard fiber is not directly related to the tangent space at each point. Dirac's monopole itself is an example of this type of quantization. In the induced case the gauge field is a 1-form $\mathbf{A}$\footnote{This can be easily generalized to higher dimensions when the gauge field $\mathbf{A}$ is a $n$-form as is shown in \cite{Nepomechie,Teitelboim}.} with components taking values in the Lie algebra $\mathfrak{g}$ of the structure group $\mathcal{G}$. 
The discrete relationship arises when we impose regularity on the gauge field so that is well defined over the entire pfb. 

The topological quantization is considered to be intrinsic when the structure group of the pfb is determined by the intrinsic symmetries of the base space. In this case, the resulting pfb is equivalent to that of the tangent bundle. Throughout this work we will be mainly interested in applying intrinsic quantization.

We will use in this work the intuitive notion of a geometric representation of a physical system that consists of a collection of geometric entities related in such a way that they describe the system. Let us assume that, as part of a geometric representation of a given physical system, there is a manifold $M$, covered by a family of open sets $U_i$, with a metric $\boldsymbol{g}$. In the tangent bundle $TM$, the connection $\tilde{\omega}$ coming from the lifting of the metric connection $\omega$ on $M$ is the one that satisfies the relation $\sigma_{i}^{\ast}\tilde{\omega}=\omega_{i}$ for any section $\sigma_{i}$ defined over $U_{i}$. To analyze the topology of $TM$, we can use $\tilde{\omega}$ and compute the characteristic class of the corresponding pfb $C(P)=C(TM)$. As we will see below, the details of how to perform this calculation will depend on the signature and dimensionality of $M$, but for the moment we just need the fact that the integral of $C(TM)$ over $M$ is bound to give a integer number \cite{damas}, i.e. 
\begin{equation}
\int C(TM)=\chi, \quad \chi \in \mathbb{Z}. \label{charclass}
\end{equation}

The information captured in the geometric representation of the physical system will be carried to $C(TM)$, so it will depend on some of the parameters, which we can call ${a_{\alpha}}$, that enter in the description of the physical system and in general it will take different values on different points of $M$. Being $\chi$ the integral of $C(TM)$ over the manifold, it will only depend on the parameters ${a_{\alpha}}$ so it can be written as a function of them, $f(a_\alpha)$. Using (\ref{charclass}) we can write

\begin{equation}\label{espectro}
f(a_{\alpha})=\chi, \quad \chi\in\mathbb{Z}.
\end{equation} 

This equation defines the discrete values that constitute the topological spectrum, giving us the chance of analyzing the physical meaning it could carry.

After the fundamentals of this formulation of topological quantization were presented and applied for the first time to gravitational configurations \cite{Patino:2002jp, Patino:2004wc}, the analysis of classical conservative systems with $n$ degrees of freedom was carried out in \cite{Nettel:2006pj, Nettel:2007zz, Nettel03}. 
Recently, the formalism of topological quantization was applied to bosonic strings in a Minkowski background \cite{Arciniega0}.

\section{Scalar fields as a harmonic map}
 
Mathematically, a scalar field $\phi$ in $m$ dimensions is a mapping $\phi:M\mapsto\mathbb{K}$, where $\mathbb{K}$ could be either $\mathbb{R}$ or $\mathbb{C}$ and $M$ is a differentiable manifold. If there are $n$ such fields, in general we can describe their physical behaviour by a variational principle applied to the action
\begin{equation}
S=\int {\cal L}(\phi^i,\partial_a\phi^i)\, \rm{d}^mx, \label{FTaction}
\end{equation}
where the index $a$ refers to some local coordinates for $M$ and $i=1,\ldots, n$.

On the other hand, we can construct a $n$-dimensional target manifold $N$ and represent the fields as a mapping $\boldsymbol{X}:M\mapsto N$ by expressing $\boldsymbol{X}$ in local coordinates $X^i$ such that each component represents one of the $n$ fields, $\phi^i \equiv X^i$. 
Noticing that the construction just described is the embedding of $M$ into $N$, the natural question arises about whether it is possible to provide $M$ and $N$ with corresponding metrics $g_{ab}$ and $G_{ij}$, such that the condition to be a critical point of (\ref{FTaction}) coincides with the condition for this embedding to be a harmonic map, that is, to be a stationary point of the Dirichlet energy functional \cite{Eells}
\begin{equation}\label{accionarmonica}
S_h=\int d^{m}x\,\sqrt{|g|}\,g^{ab}(x)\partial_{a}X^{i}\partial_{b}X^{j}G_{ij}(X),
\end{equation}
where $g=det(g_{ab})$ and $x^a$, $X^i$ are local coordinates for $M$ and $N$, respectively.

To answer this question notice that the Euler-Lagrange equations that follow from the variation of the action (\ref{accionarmonica}) with respect to $X^i$ are
\begin{equation}\label{ecmovarmonica}
\frac{1}{\sqrt{|g|}}\partial_{a}(\sqrt{|g|}g^{ab}\partial_{b}X^{i})+\Gamma^{i}_{jk}(X)\partial_{a}X^{j}\partial_{a}X^{k}g^{ab}=0,
\end{equation}
where $\Gamma^{i}_{jk}$ are the Christoffel symbols for $G_{ij}$. When all the Christoffel symbols vanish, (\ref{ecmovarmonica}) reduces to
\begin{equation}\label{kghar}
\frac{1}{\sqrt{g}}\partial_{a}(\sqrt{|g|}g^{ab}\partial_{b}X^{i})=0,
\end{equation}
which solutions are harmonic functions and hence the name of the mapping.

Here we notice that (\ref{kghar}) with $\sqrt{|g|}g^{ab} = \eta^{ab}$ are the Klein-Gordon equations for some free massless fields $X^{i}$, whereas if we were to study a different theory we would need to solve for the appropriate $\boldsymbol{g}$ and $\boldsymbol{G}$ such that (\ref{ecmovarmonica}) became the equations describing the theory of interest. This, of course, involves a intricate process of reconstructing the metric from the proper Christoffel symbols that reproduce the field equations for the theory in question.

Although this is in principle possible, instead, we will consider a harmonic map that is known to be equivalent to a system of eight free massive Klein-Gordon fields in two dimensions.
\section{Non-interacting massive bosonic field}\label{harmap}

To describe the non interacting massive bosonic fields consider a two-dimensional differentiable manifold $M$ with metric $g_{ab}$ and a ten-dimensional differentiable manifold $N$ as the target space with metric $\boldsymbol{G}$ whose components in terms of the light-cone coordinates $X^{+}=\frac{1}{\sqrt{2}}(X^{0}+X^{1})$, $X^{-}=\frac{1}{\sqrt{2}}(X^{0}-X^{1})$ and $X^{I=1,\ldots,8}$ are \cite{Blau:2001ne,Russo2004,Sadri2003}
\begin{eqnarray}\label{metG} \nonumber 
 G_{+-}=G_{-+}=-1, & G_{--}=0, & G_{++}=-\mu^{2}\sum_{I=1}^{8}X^{I}X_{I},\\ 
  G_{IJ}=G_{JI}=\delta_{IJ}, & \{ I,J \}=1,\ldots ,8. & 
\end{eqnarray}

In the light cone coordinates the explicit form of the action (\ref{accionarmonica}) is
\begin{eqnarray} \label{strtot}\nonumber
 S=-\frac{1}{4\pi \alpha'}\int \int \sqrt{-g} g^{ab}\, \left[-2\partial_{a}X^{+}\partial_{b}X^{-} +\sum_{I=1}^{8}\partial_{a}X^{I}\partial_{b}X^{I}\right. \\
 \phantom{S=-\frac{1}{4\pi \alpha'}\int \int \sqrt{-g} g^{ab}} -\left.\mu^{2}\left(\sum_{I=1}^{8}X_{I}X^{I}\right)\partial_{a}X^{+}\partial_{b}X^{+}\right] \rm{d}\sigma \rm{d}\tau.
\end{eqnarray}

We will use the light cone gauge given by
\begin{equation}\label{normadiff}
X^{+}=\alpha' p^{+}\tau, \quad p^{+}\geq 0,
\end{equation}

\noindent where $p^+$ is the initial value of the momentum along the $X^+$ direction.

Varying the action (\ref{strtot}) with respect to the metric $g^{ab}$, and demanding that it is stationary
\begin{equation}\label{constriccion}
\frac{\delta S}{\delta g^{\tau\sigma}}=0, \quad \frac{\delta S}{\delta g^{\tau\tau}}=-\frac{\delta S}{\delta g^{\sigma\sigma}}=0,
\end{equation}

\noindent we obtain the constraint equations $T_{ab} = 0$. Using the conformal gauge $\sqrt{|g|}g^{ab} = \eta^{ab}$ in the worldsheet, these are expressed as
\begin{eqnarray}\label{constrains}
&& \partial_{\sigma}X^{-}=\frac{1}{\alpha p^{+}}\sum_{I=1}^{8}\partial_{\sigma}X^{I}\partial_{\tau}X^{I}\\ 
&& \partial_{\tau}X^{-}=\frac{1}{2\alpha p^{+}}\sum_{I=1}^{8}[\partial_{\tau}X^{I}\partial_{\tau}X^{I}+\partial_{\sigma}X^{I}\partial_{\sigma}X^{I}-(\mu\alpha' p^{+})^{2}X^{I}X^{I}].
\end{eqnarray}

From the above equations we can see that $X^{-}$ is not a dynamical variable, and considering (\ref{normadiff}) we can write the action (\ref{strtot}) as
\begin{equation} \label{sboslc}
S=-  \frac{1}{4\pi \alpha'}\int \int_{0}^{2\pi\alpha' p^{+}}  \,\sum_{I=1}^{8} [-\partial_{\tau}X^{I}\partial_{\tau}X^{I}+\partial_{\sigma}X^{I}\partial_{\sigma}X^{I} + \mu^{2}\,X_{I}X^{I}]\rm{d}\sigma \, \rm{d}\tau,
\end{equation} 

\noindent where we rescaled $\tau$ and $\sigma$ by $\alpha' p^{+}$ and periodic conditions on $\sigma$ have been imposed.
Observe that (\ref{sboslc}), which represents a harmonic map, can be interpreted as the action for eight non interacting massive Klein-Gordon fields in two dimensions, by identifying $\mu$ as the mass of the field.

We also notice that there are two manifolds involved in the geometric representation, so both are in principle subject to the procedure of topological quantization.


\section{Topological spectrum of the target space}

The metric (\ref{metG}) together with the five-form 
\begin{equation}\label{5forma}
F_{+1234}=F_{+5678}=2\mu, 
\end{equation}
and a constant dilaton field
\begin{equation}\label{dilaton}
 \phi=\rm{constant}
\end{equation}
constitute a solution in Supergravity IIB \cite{Russo2004} representing a pp-wave \cite{Penroselimit}.

For this spacetime we have a ten dimensional manifold as a base space with the Lie group $SO(1,9)$ isomorphic to the standard fiber, so the invariant that must be computed is Euler's invariant. However, a quick calculation shows that Euler's form computed from the metric connection associated to (\ref{metG}) turns out to be zero, and therefore the Euler's invariant for this target space does too.
We conclude that the procedure of intrinsic topological quantization does not impose any restrictions for this part of the geometric representation.

Interestingly, the five-form (\ref{5forma}) is known to be subject to Dirac (induced) quantization \cite{Dirac:1931kp,Becker,Nepomechie,Teitelboim} and this leads to the condition
\begin{equation}\label{mu}
\mu=\phi\sqrt{(\pi/2) n},\quad n\in \mathbb{Z},
\end{equation} 
which will be used latter in this work.


\section{Topological spectra of the embedded space}

Let us now apply topological quantization to the other manifold involved in this geometric configuration. For this we construct a pfb taking the embedded manifold as the base space with metric
\begin{equation}
h_{ab}=\partial_{a}X^{i}\partial_{b}X^{j}G_{ij},\label{indmet}
\end{equation}
which after considering (\ref{constrains}) takes the following form
\begin{equation}
h_{ab}=f\eta_{ab}=f\left( \begin{array}{cc} -1 & 0 \\ 0 & 1 \end{array}\right),
\end{equation}
where $f=\sum_{I=1}^{8}\partial_{\sigma}X^{I}\partial_{\sigma}X^{I}$.

For the tangent space at each point we will use a semiorthonormal basis $\boldsymbol{e}^{\mu} = e^{\mu}_{\ a}\, d\sigma^a$, $\sigma^{a}=\{\tau,\sigma\}$, satisfying
\begin{equation}
e^{\mu}_{\ a}e^{\nu}_{\ b}\eta_{\mu\nu}=h_{ab}
\end{equation}
and, accordingly, the spin connection $\boldsymbol{\omega}$ in the standard way
\begin{equation}\label{spinchris}
(\omega^{\mu}_{\ \nu})_{a}=e^{\mu}_{\ b}\partial_{a}e^{\ b}_{\nu}+e^{\mu}_{\ b}\Gamma^{b}_{\ ac}e^{\ c}_{\nu}.
\end{equation}

Given the signature and dimensionality of $M$, the components $(\omega^{\mu\nu})$ of the one-form $\boldsymbol{\omega}$ will be elements of the Lie algebra of $SO(-1,1)$, so the characteristic class we need to compute is the Euler class $e(TM)$ given, in this case, by
\begin{equation}\label{eulerinv}
e(TM)=\frac{-1}{4\pi}\epsilon_{\ \mu}^\nu\mathbf{R}^\mu_{\ \nu},
\end{equation}
where $\epsilon_{\ \mu}^\nu$ is the Levi-Civita tensor and $\mathbf{R}^{\mu}_{\ \nu}$ is the curvature two-form $R^{\mu}_{\ \nu ab}d\sigma^{a}\wedge d\sigma^{b}$ with components	
\begin{equation}\label{curvaturadosf}
R^{\mu}_{\ \nu ab}=\partial_{a}\omega^{\mu}_{\ \nu b}-\partial_{b}\omega^{\mu}_{\ \nu a}+\omega^{\mu}_{\ \gamma a}\omega^{\gamma}_{\ \nu b}-\omega^{\mu}_{\ \gamma b}\omega^{\gamma}_{\ \nu a}.
\end{equation}
From the form of the metric (\ref{indmet}) the expression (\ref{eulerinv}) is explicitly
\begin{equation}\label{eulerexp}
 e(TM)=\frac{1}{4\pi}\left[\partial_{\sigma}\left(\frac{\partial_{\sigma}f}{f}\right)-\partial_{\tau}\left(\frac{\partial_{\tau}f}{f}\right)\right] 
 d\tau\wedge d\sigma ,
\end{equation}
so we need now to compute $f$, which requires the explicit expression for the $X^I$'s.
The equation of motion for $X^I$ is
\begin{equation}\label{ecuac}
(\partial_{\tau}^{2}-\partial_{\sigma}^{2}-\mu^{2})X^{I}=0,
\end{equation}
with the general solution, satisfying periodic conditions, given by
\begin{eqnarray}\label{solgen}\nonumber
 X^{I}=x_{0}^{I}\,\cos\mu\tau
+\frac{p_{0}^{I}}{\mu p^{+}}\sin\mu\tau
+\sqrt{\frac{\alpha'}{2}}\sum_{n=1}^{\infty}\frac{1}{\sqrt{\omega_{n}}}\left\{\alpha^{I}_{n}\exp[\frac{-\rm{i}}{\alpha' p^{+}}(\omega_{n}\tau+n\sigma)]
\right.\\ 
+\tilde{\alpha}^{I}_{n}\exp[\frac{-\rm{i}}{\alpha' p^{+}}(\omega_{n}\tau-n\sigma)]
+\left.
\alpha^{\dagger I}_{n}\exp[\frac{+\rm{i}}{\alpha' p^{+}}(\omega_{n}\tau+n\sigma)]
+\tilde{\alpha}^{\dagger I}_{n}\exp[\frac{+\rm{i}}{\alpha' p^{+}}(\omega_{n}\tau-n\sigma)]\right\},
\end{eqnarray} 
 where 
\begin{equation}\label{omegaenergia}
\omega_{n}=\sqrt{n^{2}+(\mu\alpha' p^{+})^{2}}, \quad n\in\mathbb{N},
\end{equation}
and $\alpha_n$, $\tilde{\alpha}_{n}$ are the coefficients of the $n$-th right and left modes of oscillation, respectively.

From here we see that the values we assign to the $\alpha$'s and $\tilde{\alpha}$'s will determine $e(TM)$, and therefore its integral over $M$
\begin{equation}\label{eulerclas}
\int_{M} e(TM)=\chi_{TM}, \quad 
\end{equation}
will be a function $\chi_{TM}(\alpha_n,\tilde{\alpha}_n)$ of these coefficients and the frequencies.

Given that in general $M$ is not compact, we will have to cut a finite part of it and add the contribution that the so induced boundaries have to the invariant we are interested in. So we will need to treat surfaces $M_F$ with boundary, for which the invariant we are looking for is given by
\begin{equation}
\chi_{TM}=\int_{M_F} e(TM) + \frac{1}{2\pi}\int_{\partial M_F} \kappa ds, \label{invarcomple}
\end{equation}
where $\kappa$ is the geodesic curvature of the boundary and $ds$ is the proper arch length along it.

From the reasons exposed above, on the one hand, $\chi_{TM}$ will be a function of the $\alpha$'s and $\tilde{\alpha}$'s, and on the other hand it will be bound to be an integer, {\it i.e.} $\chi_{TM}(\alpha_n,\tilde{\alpha}_n)\in\mathbb{Z}$. This last condition is the one that determines the topological spectrum for the physical parameters $\alpha_n,\tilde{\alpha}_n$ and $\omega_n$.



\section{Topological spectrum of particular configurations}\label{sec:emb}

Now, we can proceed to compute the explicit form of $\chi_{TM}(\alpha_n,\tilde{\alpha}_{n})\in\mathbb{Z}$, extract the topological spectrum and analyze its physical content. 

It would seem natural to try to integrate (\ref{eulerexp}) over $M$ in its more general form to get the topological spectrum; however, by doing so we would have to manage infinite sums over the modes of oscillation making the integration a complicated technical problem. In order to reach concrete expressions for the topological spectra we can consider particular configurations for the fields instead. 

We can start by considering the solutions (\ref{solgen}) with the fewest number of coefficients different from zero that still give a non trivial result for $\chi_{TM}(\alpha_n,\tilde{\alpha}_{n})$, and extract a condition from each of these cases, which will be the topological spectrum. 

The simplest solution leading to a non trivial Euler invariant is given by (\ref{solgen}) with only $\alpha_1^1$ and $\tilde \alpha_{1}^2$ different from zero, so all the fields $X^{i\neq \{ 1,2\} }$ vanish identically, or  they describe, at most, the motion of the center of mass $X^{i}=x_{0}^{i}\,\cos(\mu\tau)+\frac{p_{0}^{i}}{\mu p^{+}}\sin(\mu\tau)$, while for $X^{1}$ and $X^{2}$ we have
\begin{eqnarray}\label{firstsol}
X^{1}&=&x_{0}^{1}\,\cos(\mu\tau)+\frac{p_{0}^{1}}{\mu p^{+}}\sin(\mu\tau)+\sqrt{\frac{\alpha'}{2}}\frac{2r}{\sqrt{\omega_1}}\cos \left[\frac{1}{\alpha'p^{+}}(\omega_1\tau+\sigma)+\gamma \right]\\
X^{2}&=&x_{0}^{2}\,\cos(\mu\tau)+\frac{p_{0}^{2}}{\mu p^{+}}\sin(\mu\tau)+\sqrt{\frac{\alpha'}{2}}\frac{2\tilde{r}}{\sqrt{\omega_1}}\cos \left[\frac{1}{\alpha'p^{+}}(\omega_1\tau-\sigma)+\tilde{\gamma}\right],
\end{eqnarray}
where we wrote  $\alpha_1^1$ and $\tilde{\alpha}^2_{1}$ in their polar representation $r e^{-i\gamma}$ and $\tilde{r} e^{-i\tilde{\gamma}}$, respectively.

The Euler form for this field configuration is
\begin{eqnarray}\label{eulersigmatau}\nonumber
e(M)=\left\{r^{2}\tilde{r}^{2}\left[ (\omega^{2}_{1}-1) \cos2\left( \frac{\omega_1\tau+\sigma}{\alpha'p^{+}}+\gamma\right)-\cos2\left( \frac{\omega_1\tau-\sigma}{\alpha'p^{+}}+\tilde{\gamma}\right)\right.\right.\\ \nonumber
+\omega^{2}_{1}\cos2\left( \frac{\omega_1\tau-\sigma}{\alpha'p^{+}}+\tilde{\gamma}\right)-2\omega^{2}_{1}\cos\left(2(\gamma-\tilde{\gamma})+\frac{4\sigma}{\alpha'p^{+}}\right)\\ \nonumber
+\left.2\cos\left(2(\gamma+\tilde{\gamma})+\frac{4\omega_{1}\tau}{\alpha'p^{+}}\right)\right] -2r^{4}(\omega^{2}_{1}-1)\sin^{2}\left(\frac{\omega_1\tau+\sigma}{\alpha'p^{+}}+\gamma\right)\\ \nonumber
\left.-2\tilde{r}^{4}(\omega^{2}_{1}-1)\sin^{2}\left(\frac{\omega_1\tau-\sigma}{\alpha'p^{+}}+\tilde{\gamma}\right)\right\} /\left\{4\pi (\alpha'p^{+})^{2} \left[r^{2}\sin^{2}\left(\frac{\omega_1\tau+\sigma}{\alpha'p^{+}}+\gamma\right)\right.\right.\\
+\left.\left.\tilde{r}^{2}\sin^{2}\left(\frac{\omega_1\tau-\sigma}{\alpha'p^{+}}+\tilde{\gamma}\right)\right]^{2}\right\} \rm{d}\tau \wedge \rm{d}\sigma. 
\end{eqnarray}

To compute $\chi_{TM}$ without worrying about possible boundary terms generated by considering a finite part $M_F$ of $M$, we notice that the induce metric is periodic in $\tau$, and so it will be $\kappa$. If we consider $M_F$ to cover an integer number of periods, the contribution of the boundary at one end in $\tau$ will exactly cancel the contribution at the other extreme, since the values of $\kappa$ will be the same at both ends, but the orientation of the boundary will be reverse, giving and overall minus sign. If we consider $M_F$ to cover exactly one period, the topological spectrum will be given by the condition
\begin{equation}\label{eulerclasfini}
\int_{M_F} e(TM)\in\mathbb{Z}. 
\end{equation}

To integrate this expression over $M_F$ we change variables to
\begin{equation}\label{cambiovar}
x=\sin\left( \frac{\omega_1\tau+\sigma}{\alpha'p^{+}}+\gamma\right) \quad \mathrm{and} \quad
y=\sin\left( \frac{\omega_1\tau-\sigma}{\alpha'p^{+}}+\tilde{\gamma}\right),
\end{equation}
in terms of which the Euler form acquires the form
\begin{eqnarray}\nonumber
e(TM)=\left(\frac{r^{4}x^{2}(\omega_1^{2}-1)+\tilde{r}^{4}y^{2}(\omega_1^{2}-1) 
+r^{2}\tilde{r}^{2}x^{2}(4y^{2}-1)(\omega_1^{2}-1)}{4\pi\sqrt{1-x^{2}}\sqrt{1-y^{2}}\omega_1(r^{2}x^{2}+\tilde{r}^{2}y^{2})^{2}}\right.\\
-\left.\frac{r^{2}\tilde{r}^{2}[y^{2}(\omega_1^{2}-1)+4xy\sqrt{1-x^{2}}\sqrt{1-y^{2}}(\omega_1^{2}+1)]}{4\pi\sqrt{1-x^{2}}\sqrt{1-y^{2}}\omega_1(r^{2}x^{2}+\tilde{r}^{2}y^{2})^{2}}\right) \rm{d} x \wedge \rm{d} y.
\end{eqnarray}

It turns out that the outcome of the integral depends on the order of integration. In fact,  
integrating first with respect to $x$ and then with respect to $y$ we get
\begin{equation}  \label{eul1}
\int_{-1}^{1} \int_{-1}^{1}e(TM)dx\,dy=\frac{(1-\omega_{1}^{2})}{2\,\omega_{1}}\frac{r}{\tilde{r}},
\end{equation}
while in the inverse integration order the result is
\begin{equation}   \label{eul2}
\int_{-1}^{1} \int_{-1}^{1}e(TM)dy\,dx=\frac{(1-\omega_{1}^{2})}{2\,\omega_{1}}\frac{\tilde{r}}{r}.
\end{equation}
This non-commutativity in the order of integration is due to the presence of singular points within the domain of integration. To guarantee the existence of such integral it is necessary to define a neighborhood $B_\epsilon(p_0)$ of radius $\epsilon$ around the singular point $p_0$, compute the integral over $M_F - B_\epsilon(p_0)$ adding the contribution due to the integral of $\kappa$ over the boundary $\partial B_\epsilon(p_0)$ of $B_\epsilon(p_0)$ and take the limit $\epsilon\rightarrow 0$.

\begin{figure}[htb]
\begin{center}
\includegraphics[width=10cm]{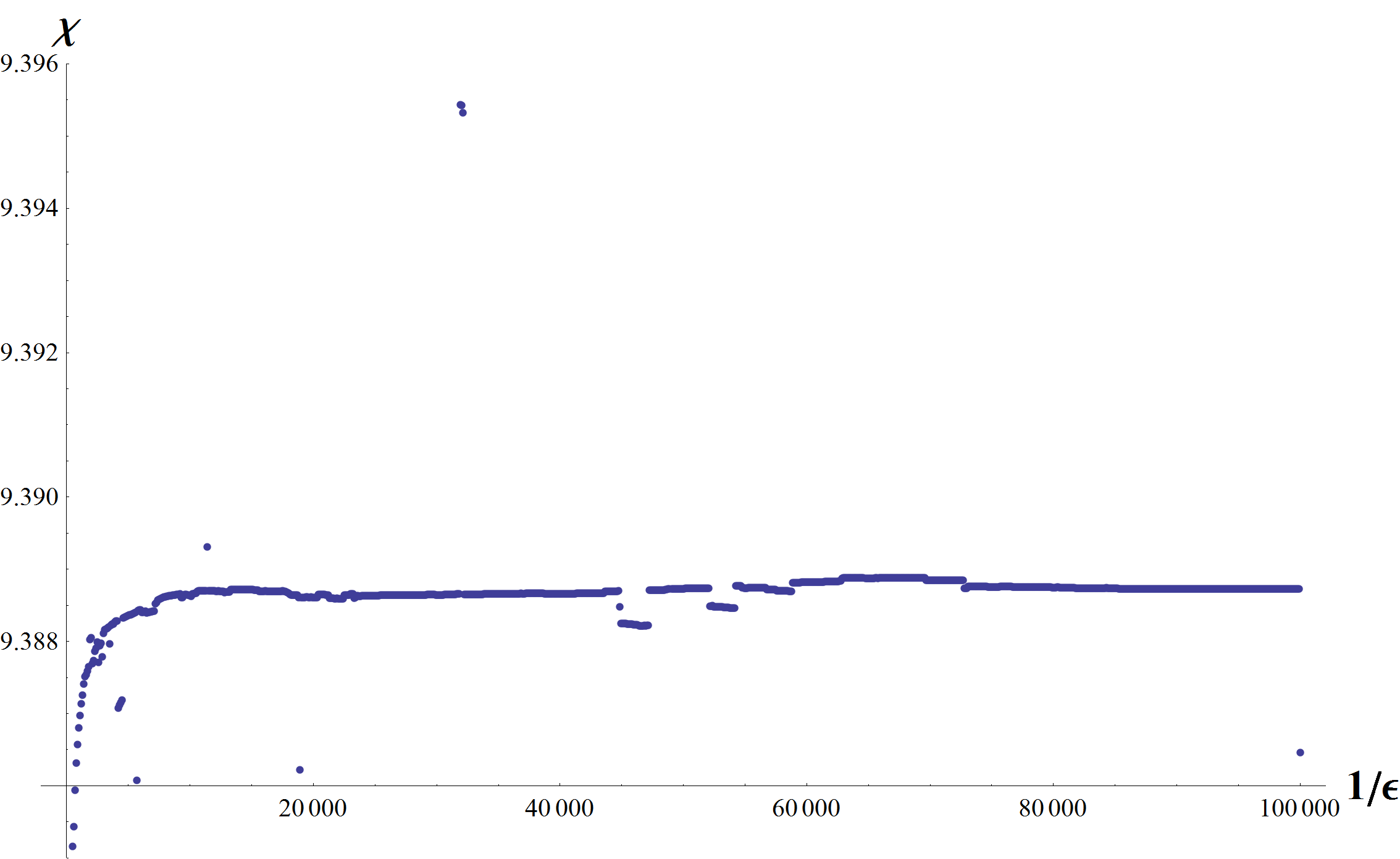}
\caption{\small{Plot of the Euler invariant $N[\chi_{TM}](r,\tilde{r})$ computed using the numerical method described in the text  taking $r=\tilde{r}=1$, $\omega_{1}=2$, $p^{+}=1$ and $\alpha'=1$}}
\label{eulerNumerico}
\end{center}
\end{figure}

\begin{figure}[htb]
\begin{center}
\includegraphics[width=10cm]{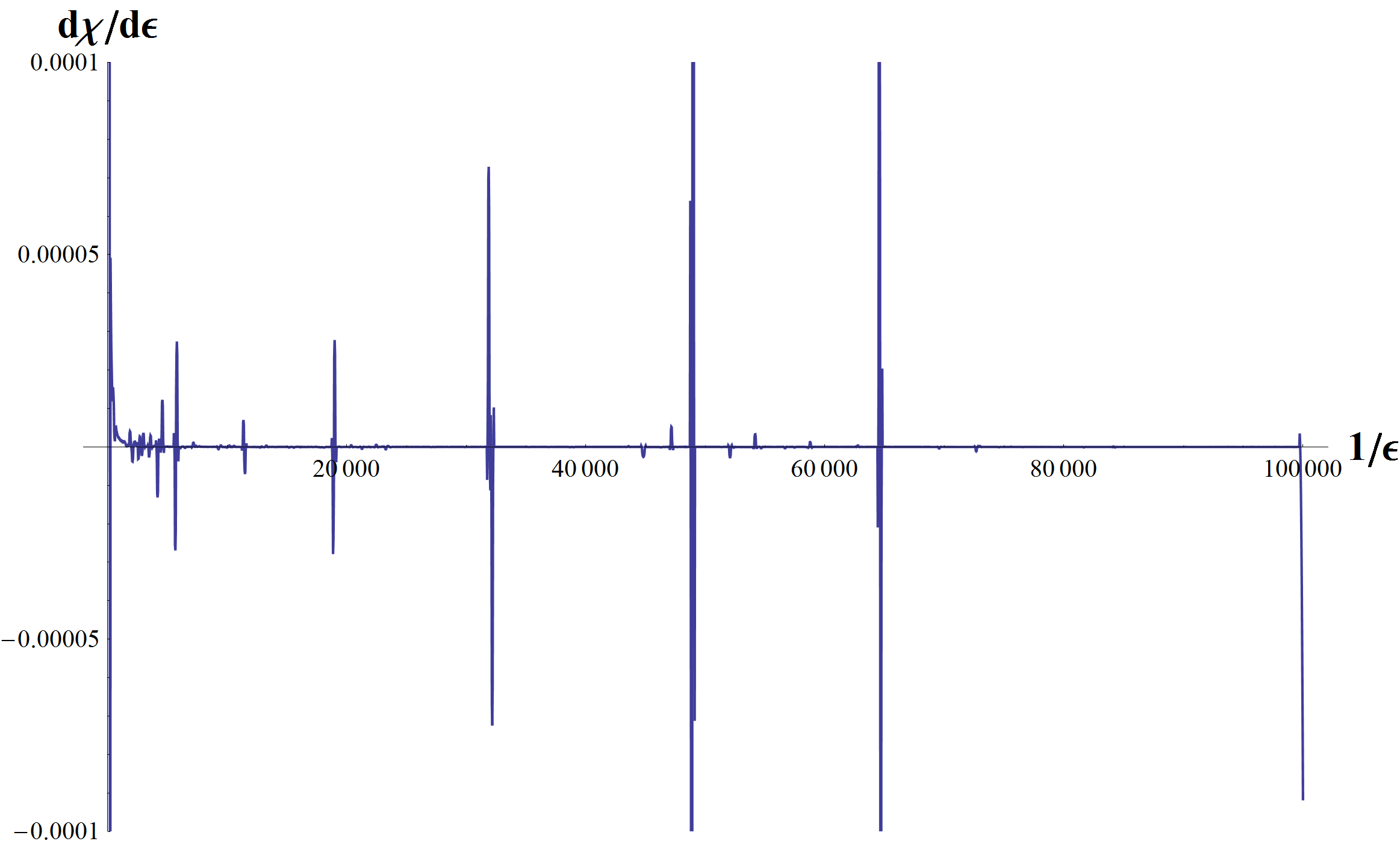}
\caption{\small{Plot of the derivative of the Euler invariant $N[\chi_{TM}](r,\tilde{r}$) as a function of $\epsilon$  taking $r=\tilde{r}=1$, $\omega_{1}=2$, $p^{+}=1$ and $\alpha'=1$}}
\label{derivada}
\end{center}
\end{figure}

After following such a procedure we find that the invariant is given by the mean value of the two Euler characteristic (\ref{eul1}, \ref{eul2}) 
\begin{equation}  \label{eulerbos}
\chi_{TM}= \frac{1}{2}\left(\int_{M}e(TM)dx dy + \int_{M}e(TM)dy dx \right)   = \frac{(\alpha' p^{+}\mu)^{2}}{4\,\omega_{1}}\left(\frac{r}{\tilde{r}}+\frac{\tilde{r}}{r}\right).
\end{equation}
As a consistency check for (\ref{eulerbos}), we also performed both numerical integrations, the one over $M_F$ and the one over $\partial B_\epsilon(p_0)$, and add them. To verify the convergence of the integral, this computation was performed for a set of fixed values of $r$ and $\tilde{r}$ with $\epsilon$ ranging from 0.01 to $10^{-6}$, satisfactory confirming the existence and finiteness of the integral. As an example of this verification process we show in figure (\ref{eulerNumerico}) the value of the sum of the integrals as a function of $\epsilon$ for $r=\tilde{r}=1$. Figure (\ref{derivada}) depicts the derivative with respect to $\epsilon$ of the function shown in figure (\ref{eulerNumerico}). 

\begin{figure}[htb]
\begin{center}
\includegraphics[width=15cm]{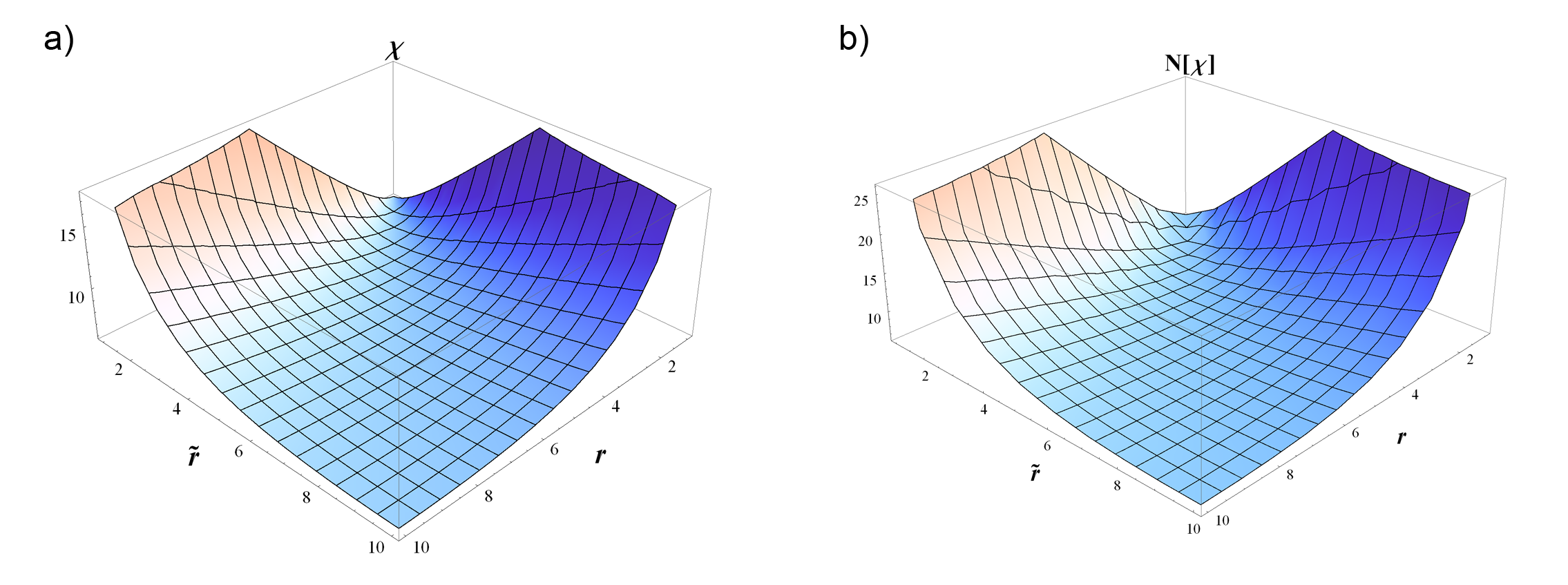}
\caption{\small{Plot of the Euler invariant $\chi_{TM}(r,\tilde{r}$) computed using {\bf a)} expression (\ref{eulerbos}) and {\bf b)} the numerical method described in the text  taking $\omega_{1}=2$, $p^{+}=1$ and $\alpha'=1$}}
\label{espectro2}
\end{center}
\end{figure}

In Fig. \ref{espectro2} we plot $\chi_{TM}$ as a function of $r$ and $\tilde{r}$ computed by (a) using (\ref{eulerbos}) and (b) the numerical method for $\epsilon=0.001$, which is the value at which, for a good range of values for $r$ and $\tilde{r}$, the integral does not seam to vary much as we decrees $\epsilon$.

We can see the behavior to be very similar in these two plots except for any numerical error introduced by $\epsilon\neq 0$.

Now we are ready to use the Gauss-Bonnet theorem applied to (\ref{eulerbos}), then establishing that $\chi_{TM}=k\in \mathbb{Z}$ and obtaining the quantization condition which can be written as
\begin{equation}
r=\frac{1}{(\alpha' p^{+}\mu)^{2}}\left(2\omega_{1}k\pm\sqrt{(2\omega_{1}k)^{2}-(\alpha' p^{+}\mu)^{4}}\right)\tilde{r}. \label{spect1}
\end{equation}
We see that for a given mass $\mu$ of the bosonic field the ratio of the amplitudes $r/\tilde r$ is not arbitrary but can have only discrete values which depend on the integer $k$.

To move forward we can consider a pair of arbitrary modes for this particular configuration of a left and a right mode for the fields $X^1$ and $X^2$, respectively. Then, we take the non zero coefficients in (\ref{solgen}) to be $\alpha_m^1$ and $\tilde{\alpha}_n^2$ so that the relevant fields are given by  
\begin{eqnarray}\label{2campos}
X^{1}&=&x_{0}^{1}\,\cos(\mu\tau)+\frac{p_{0}^{1}}{\mu p^{+}}\sin(\mu\tau)+\sqrt{\frac{\alpha'}{2}}\frac{2r_m}{\sqrt{\omega_{m}}}\cos[\frac{1}{\alpha'p^{+}}(\omega_{m}\tau+m\sigma)+\gamma_m],\\
X^{2}&=&x_{0}^{2}\,\cos(\mu\tau)+\frac{p_{0}^{2}}{\mu p^{+}}\sin(\mu\tau)+\sqrt{\frac{\alpha'}{2}}\frac{2\tilde{r}_n}{\sqrt{\omega_{n}}}\cos[\frac{1}{\alpha'p^{+}}(\omega_{n}\tau-n\sigma)+\tilde{\gamma}_n].
\end{eqnarray}

\begin{figure}[htb]
\begin{center}
\includegraphics[width=6cm]{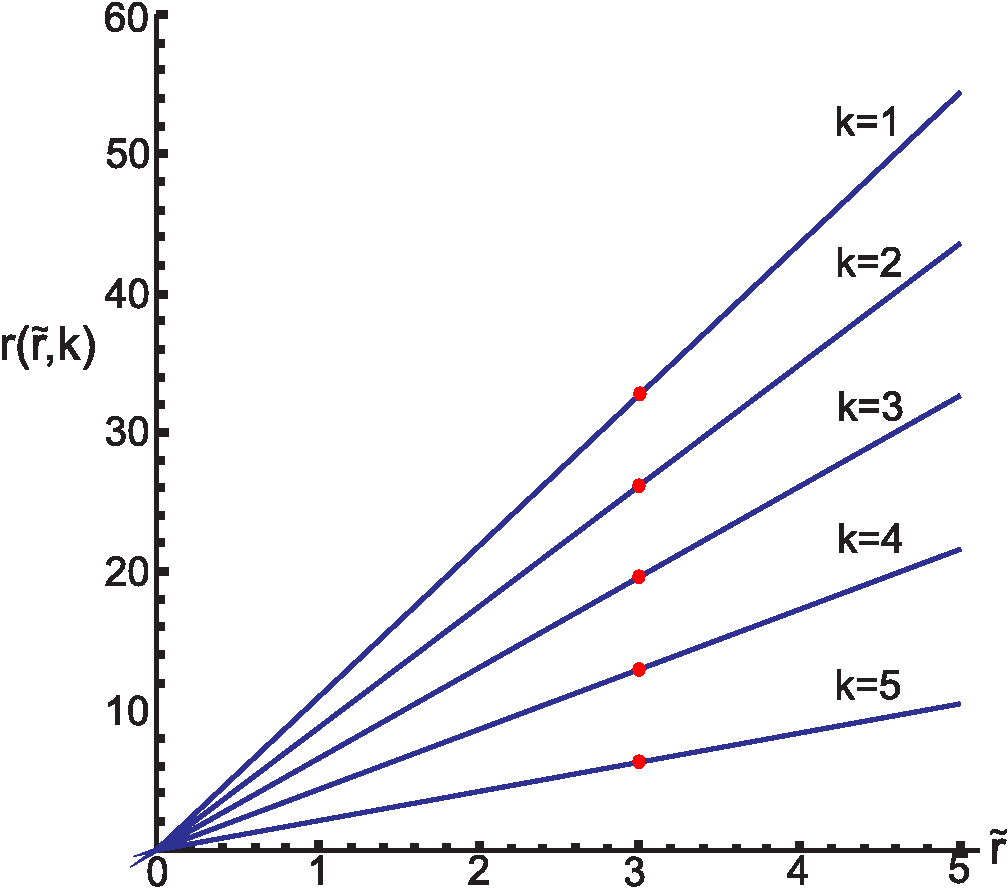}
\caption{\small{Plot of equation (\ref{condicion2campos}). Each line is for $n=1$, $m=3$, $\mu=2$, $\alpha'=1$, $p^{+}=1$ and $k=1,\ldots,5$. The red dots indicate the posible values of $r_{m}$ for each $\tilde{r}_{n}\in \mathbb{N}$ fixed. }}
\label{rr2}
\end{center}
\end{figure}

Then, following the definition of topological spectrum given in (\ref{eulerbos}) we obtain 
\begin{equation}\label{resultado1b}
 \chi_{TM}= \frac{(\alpha' p^{+}\mu)^2}{2(n\omega_{m}+m\omega_{n})}\left(\frac{m\sqrt{\omega_{n}}r_{m}}{n\sqrt{\omega_{m}}\tilde{r}_{n}}+\frac{n\sqrt{\omega_{m}}\tilde{r}_{n}}{m\sqrt{\omega_{n}}r_{m}} \right) = k,
\end{equation}
where $k \in \mathbb{Z}$. The quantization condition can be expressed as
\begin{equation}\label{condicion2campos}
r_{m}=\frac{1}{(\alpha' p^{+}\mu)^{2}}\left[(n\omega_{m}+m\omega_{n})k\pm\sqrt{(n\omega_{m}+m\omega_{n})^{2}k^{2}-(\alpha' p^{+}\mu)^{4}}\right]\left(\frac{n\sqrt{\omega_{m}}}{m\sqrt{\omega_{n}}}\right)\tilde{r}_{n}.
\end{equation}
This topological spectrum generalizes (\ref{spect1}) and tells us that the ratio of amplitude modes $r_{m}/\tilde{r}_{n}$ fills for each $m$ and $n$ a discrete spectrum whose states are labeled by $k$. We can also conclude that if we fix the value of one of the amplitude modes, say $\tilde{r}_{n}$, then $r_{m}$ can take only a discrete set of values. This discrete behavior is illustrated in Fig. \ref{rr2}.

Yet another case we can analyze is when there are two modes with amplitudes $\alpha^{1}_{m}$ and $\tilde{\alpha}^{1}_{n}$ excited over a single field $X^{1}$ 
\begin{eqnarray}\label{1campo}\nonumber
X^{1}=x_{0}^{1}\,\cos(\mu\tau)+\frac{p_{0}^{1}}{\mu p^{+}}\sin(\mu\tau)+\sqrt{\frac{\alpha'}{2}}\frac{2r_m}{\sqrt{\omega_{m}}}\cos \left[\frac{1}{\alpha'p^{+}}(\omega_{m}\tau+m\sigma)+\gamma_m \right]\\
+\sqrt{\frac{\alpha'}{2}}\frac{2\tilde{r}_n}{\sqrt{\omega_{n}}}\cos \left[\frac{1}{\alpha'p^{+}}(\omega_{n}\tau-n\sigma)+\tilde{\gamma}_n\right],
\end{eqnarray}
while, as in the previous case, the other fields $X^{i\neq 1}$ describe at most the motion of the center of mass.
The computation of the corresponding Euler form can be carried out straightforward and leads to 
\begin{eqnarray}\label{eulerimaginario}\nonumber
\chi_{TM}=\frac{4(nm+\omega_{m}\omega_{n})}{\pi(n\omega_{m}+m\omega_{n})}\left[\mathrm{ArcTanh}\left(\frac{m\sqrt{\omega_{n}}r_{m}}{n\sqrt{\omega_{m}}\tilde{r}_{n}}\right)+\mathrm{ArcTanh}\left(\frac{n\sqrt{\omega_{m}}\tilde{r}_{n}}{m\sqrt{\omega_{n}}r_{m}}\right)\right]\\  
+\frac{\rm{i}}{(n\omega_{m}+m\omega_{n})}\left[(\omega_{m}^{2}-m^{2})\frac{m\sqrt{\omega_{n}}r_{m}}{n\sqrt{\omega_{m}}\tilde{r}_{n}} +(\omega_{n}^{2}-n^{2})\frac{n\sqrt{\omega_{m}}\tilde{r}_{n}}{m\sqrt{\omega_{n}}r_{m}}\right] = k,
\end{eqnarray}
\noindent where $k \in \mathbb{Z}$.
Notice that in this case the result of the integration is a complex function. Using the relationship\footnote{For the sake of concreteness we consider the case $m\sqrt{\omega_{n}}r>n\sqrt{\omega_{m}}\tilde{r}$. In fact, the opposite case is equivalent because $\mathrm{ArcTanh}(1/x)=\mathrm{ArcTanh}(x)-i\pi/2$ when $ x<1$.}
\begin{equation}
\mathrm{ArcTanh}\left(\frac{n\sqrt{\omega_{m}}\tilde{r}_{n}}{m\sqrt{\omega_{n}}r_{m}}\right)=\mathrm{ArcTanh}\left(\frac{m\sqrt{\omega_{n}}r_{m}}{n\sqrt{\omega_{m}}\tilde{r}_{n}}\right)-i\frac{\pi}{2},
\end{equation}
\noindent where we considered $m\sqrt{\omega_{n}}r_{m}<n\sqrt{\omega_{m}}\tilde{r}_{n}$. Demanding that the imaginary part of (\ref{eulerimaginario}) vanishes, $\mathrm{Im}(\chi_{TM}) = 0$, we obtain the condition
\begin{equation}  \label{r*A}
r_* \equiv \frac{m\sqrt{\omega_n}r_m}{n\sqrt{\omega_m}\tilde{r}_n}  = \frac{(nm+\omega_{m}\omega_{n})\pm \sqrt{(nm+\omega_{m}\omega_{n})^{2}-(\mu\alpha'p^{+})^{4}}}{(\mu\alpha'p^{+})^{2}}.
\end{equation}
%
Then, the real part of (\ref{eulerimaginario}) must fulfill the condition $\mathrm{Re}(\chi_{TM}) = k$ yielding
\begin{equation}\label{espectropm*}
 \frac{8(nm+\omega_{m}\omega_{n})}{\pi(n\omega_{m}+m\omega_{n})}\mathrm{ArcTanh}\left(r_*  \right)=k.
\end{equation}
So we obtain an implicit discrete relationship for $r_m$ and $\tilde{r}_n$ from (\ref{r*A}-\ref{espectropm*}) that can be stated as
$r_* = r_*(k)$, i.e., $r_m = r_m(\tilde{r}_n, k)$, or explicitly
\begin{equation}
r_* = \frac{1}{(\alpha' p^+ \mu )^2}\left[ \frac{\pi}{8}\frac{(n\omega_m + m \omega_n)}{\mathrm{ArcTanh}(r_*)}k \pm \sqrt{\left(\frac{\pi}{8}\frac{(n\omega_m + m \omega_n)}{\mathrm{ArcTanh}(r_*)}\right)^2 - (\alpha' p^+ \mu)^4 }\,\right] \ .
\end{equation}

It is important to mention that we can always find $p^{+}$ such that the condition (\ref{r*A}) is satisfied for fixed values of $n$, $m$, $\mu$ and the integer $k$.


\subsection{Discretization of the energy}

Here we analyze the effect of the topological spectrum on the Hamiltonian for the particular configuration given in (\ref{firstsol}) for the free massive bosonic field. From the Lagrangian density in the action integral (\ref{sboslc})
\begin{equation}
\mathcal{L} = -\frac{1}{4\pi \alpha'} \sum_{I=1}^{8} [-\partial_{\tau}X^{I}\partial_{\tau}X^{I}+\partial_{\sigma}X^{I}\partial_{\sigma}X^{I} + \mu^{2}\,(X^{I})^2],  \nonumber
\end{equation}

\noindent the Hamiltonian density may be found in the usual way $\mathcal{H} = \sum_I \partial_\tau X^I \Pi_I - \mathcal{L}$, where the conjugate momentum is given by $\Pi_I = \frac{\delta \mathcal{L}}{\delta(\partial_\tau X^I)} = \frac{1}{2\pi \alpha'} \partial_\tau X^J \delta_{JI}$. Thus, the Hamiltonian density in general has the following form depending on the solutions for the fields $X^I$
\begin{equation}
\mathcal{H} = \frac{1}{4\pi\alpha'} \sum_I \left[\partial_\tau X^I \partial_\tau X^I + \partial_\sigma X^I \partial_\sigma X^I + \mu^2 (X^I)^2 \right],  \nonumber
\end{equation}

\noindent and the Hamiltonian can be found by integrating the Hamiltonian density over the coordinate $\sigma$, $H= \int d\sigma \mathcal{H}$. The Hamiltonian for the particular solution  considered (\ref{firstsol}) can be obtained after a simple calculation
\begin{equation}\label{hamiltoniano1}
H=\omega_{m}r_{m}^{2}+\omega_{n}\tilde{r}_{n}^{2},
\end{equation}

\noindent which, upon the imposition of the topological spectrum previously found (\ref{condicion2campos}) takes the following form
\begin{equation}
H = \left[\omega_n + \omega_m f_{m,n}^2(k) \right] \tilde{r}_n^2,
\end{equation}

\noindent where
\begin{equation}
f_{m,n}(k) =  \frac{1}{(\alpha' p^+ \mu )^2} \left[(n\omega_m + m\omega_n)k + \sqrt{(n\omega_m + m\omega_n)^2k^2 - (\alpha' p^+ \mu)^4 } \right] \frac{n\sqrt{\omega_m}}{m\sqrt{\omega_n}}.
\end{equation}

\begin{figure}[htb]
\begin{center}
\includegraphics[scale=1]{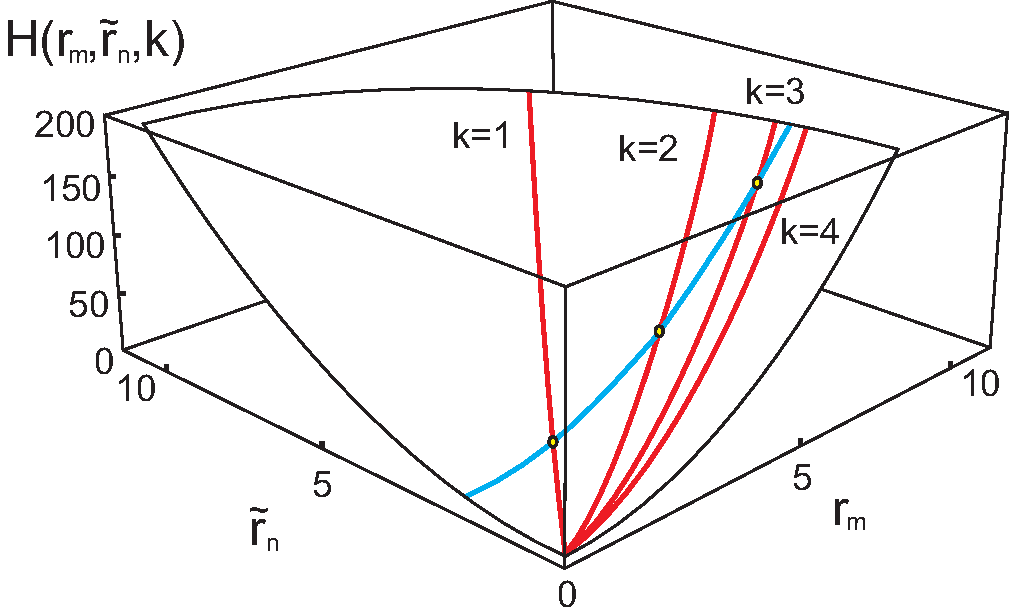}
\caption{\small{Plot of Hamiltonian for two fields (\ref{2campos}) when $n=1$, $m=3$, $\mu=2$, $\alpha'=1$ and $p^{+}=1$. The lines that intersect the origin are the condition (\ref{condicion2campos}) for the Hamiltonian when $k=1,\ldots,4$ and the line that intersect these conditions is for $\tilde{r}_{n}=2$. The dots are the allowed values of the Hamiltonian when $k$ runs over $\mathbb{Z}$. }}
\label{hrk}
\end{center}
\end{figure}

Then, for a fixed value of $\tilde{r}_n$ the Hamiltonian takes discrete values depending on the integer $k$, hence the energy of this particular configuration of the free massive bosonic field becomes discrete as it is shown in Fig.  \ref{hrk}.


\section{Discussion}
\label{sec:dis}

The topological spectra we have computed in the present work for the massive scalar field represents discrete relationships between the particular parameters that determine the field configurations, constituting the first example of topological quantization for scalar fields. These relationships depend on the number of modes considered for the scalar field description. Then, it is interesting to find out what characteristics of the field configuration result affected by this discreteness. To this end, we computed the Hamiltonian that represents the total energy for all the massive scalar field configurations discussed in this paper and found a case showing discrete behavior once one of the parameters has been fixed.


Furthermore, the topological spectra we obtained in this work can be interpreted as discrete relationships between the parameters $r_m$ and $\tilde r _n$. So we can conclude that, in general, a topological spectrum leads to a discretization of the total energy of the system. We obtained explicitly this discretization in the case 
of two fields (\ref{2campos}) for which the Hamiltonian in terms of the topological spectrum is given in Eq.(\ref{condicion2campos}). Similar results can be obtained for the other configurations investigated here.
In fact, for the case of one field with two excited modes (\ref{1campo}), we obtain a similar expression for the Hamiltonian (\ref{hamiltoniano1}) with the proper amplitudes and frequencies corresponding to the case in turn. But for this configuration, the discretization of the energy is given through $p^{+}(k)$ in (\ref{r*A}), thus we have again that $H=H(r_{m},\tilde{r}_{n},k)$ obeys a discrete behavior. 
The fact that $p^{+}$ has a spectrum is an interesting feature that, until now, only happen in this formalism.
This results means that the Hamiltonian belonging to a free massive bosonic field has a discrete behavior as a function of $k\in\mathbb{Z}$.

We found an interesting feature when integrating the Euler invariant, namely, that the integration does not commute under the interchange of the integration variables. This means that we can see the integrals
\begin{equation}
 \int (\,\,\,) \rm{d} x\equiv F_{x} \qquad \mathrm{and}\qquad \int (\,\,\, ) \rm{d} y\equiv G_{y}
\end{equation}
as operators such that obey the relation
\begin{equation}
 F_{x}G_{y}-G_{y}F_{x}\neq 0\ .
\end{equation}
This is more than a curious characteristic because $F_{x}$ and $G_{y}$ are functions of $\alpha$ and $\tilde{\alpha}$ so that, in a certain way, we found in the  topology of the system the non commuting behavior that in standard quantum field theory is only given through the imposition of this relation to the fields and its conjugate moment. A more detailed analysis will be necessary to clarify this interesting result of topological quantization.

{Acknowledgments}
This work was partially supported by DGAPA-UNAM No. IN106110 and No. IN117012-3 and CONACYT grant No. 166391.
F. N. acknowledges support from DGAPA-UNAM (postdoctoral fellowship). 

\section*{References}
\bibliographystyle{unsrt}
\bibliography{articulos}
\end{document}